\documentclass[10pt]{article}

\usepackage[english]{babel} 
\usepackage{amsmath}                
\usepackage{amsfonts}               
\usepackage{amssymb}                
\usepackage{amsopn}                 
\allowdisplaybreaks[3] 
\usepackage{amsthm}                
\usepackage{bbm}                    
\usepackage{mathrsfs}               
\usepackage[retainorgcmds]{IEEEtrantools} 
\usepackage{calc}                   
\usepackage[dvips]{graphicx}        
\usepackage{epsfig}                
\usepackage{psfrag}                 
\usepackage[dvips]{color}           
\usepackage{fancyhdr}               
\usepackage{verbatim}               
\usepackage{exscale}                
\usepackage{macros}
\newtheorem{theorem}{Theorem}

\newtheorem{note}{Note}

\newcommand{\PiG}{\Pi_{\textnormal{G}}}

\title{Gaussian Fading Is the Worst Fading}

\author{Tobias Koch ~~ Amos Lapidoth\\\small ETH Zurich\\\small Zurich, Switzerland\\
   \small Email: \{tkoch, lapidoth\}@isi.ee.ethz.ch
 }

\date{}

\begin{document}

\maketitle

\begin{abstract}
  \renewcommand{\thefootnote}{}
  The capacity of peak-power limited, single-antenna, noncoherent,
  flat-fading channels with memory is considered. The emphasis is on
  the capacity pre-log, i.e., on the limiting ratio of channel capacity to
  the logarithm of the signal-to-noise ratio (SNR), as the SNR tends
  to infinity. It is shown that, among all stationary \& ergodic
  fading processes of a given spectral distribution function and whose law
  has no mass point at zero, the Gaussian process gives rise to the
  smallest pre-log. The assumption that the law of the fading process has no mass
  point at zero is essential in the sense that there exist stationary
  \& ergodic fading processes whose law has a mass point at zero and
  that give rise to a smaller pre-log than the Gaussian process of
  equal spectral distribution function. An extension of our results
  to multiple-input single-output fading channels with memory is also
  presented.
  \footnote{The material in this paper
  was presented in part at the 2006 IEEE International Symposium on Information
  Theory (ISIT) in Seattle, Washington, USA.}
\end{abstract}
\setcounter{footnote}{0}

\section{Introduction}
\label{sec:intro}

We study the capacity of peak-power limited,
single-antenna, discrete-time, flat-fading channels with
memory. A noncoherent channel model is considered where the
transmitter and receiver are both
aware of the law of the fading process, but not of its realization.
Our focus is on the capacity at high signal-to-noise ratio
(SNR). Specifically, we study the capacity pre-log, which is defined
as the limiting ratio of channel capacity to the logarithm of the SNR,
as the SNR tends to infinity.

The capacity pre-log of \emph{Gaussian} fading
channels was derived in \cite{lapidoth05} (see also
\cite{lapidoth03_2}). It was shown that the pre-log is
given by the Lebesgue measure of the set of harmonics
where the derivative of the spectral distribution function
that characterizes the memory of the fading process is zero. To the best
of our knowledge, the capacity pre-log of \emph{non-Gaussian} fading
channels is unknown.

In this work, we demonstrate that the Gaussian assumption in the
analysis of fading channels at high SNR is conservative in the sense that for
a large class of fading processes the Gaussian process is the worst.
More precisely, we show that among all stationary \& ergodic
fading processes of a given spectral distribution function and whose law has
no mass point at zero, the Gaussian process gives rise to the smallest
pre-log.

This paper is organized as follows. Section~\ref{sec:channel}
describes the channel model. Section~\ref{sec:capacity} defines
channel capacity and the capacity pre-log. Section~\ref{sec:results}
presents our main results. Section~\ref{sec:proofprelog} provides the
proofs of these results. Section~\ref{sec:MISO} discusses the
extension of our results to multiple-input single-output (MISO) fading
channels with memory. Section~\ref{sec:discussion} concludes the paper
with a summary and a discussion of our results.

\section{Channel Model}
\label{sec:channel}
Let $\Complex$ and $\Integers$ denote the set of complex numbers and
the set of integers. We consider a single-antenna flat-fading channel with memory where the
time-$k$ channel output $Y_k \in \Complex$ corresponding to the
time-$k$ channel input $x_k \in \Complex$ is given by
\begin{equation}
  \label{eq:channel}
  Y_k = H_k x_k + Z_k, \qquad k\in\Integers.
\end{equation}
Here the random processes $\{Z_k,\,k\in\Integers\}$ and
$\{H_k,\,k\in\Integers\}$ take value in $\Complex$ and model the
additive and multiplicative noises, respectively.  It is assumed that
these processes are statistically independent and of a joint law that
does not depend on the input sequence $\{x_k\}$. 

The additive noise $\{Z_k,\,k\in\Integers\}$ is a sequence of independent and
identically distributed (IID) zero-mean, variance-$\sigma^2$,
circularly-symmetric, complex Gaussian random variables. The
multiplicative noise (``fading'') $\{H_k,\,k\in\Integers\}$ is a mean-$d$,
unit-variance, stationary \& ergodic stochastic process of spectral
distribution function $F(\lambda)$, $-1/2 \leq \lambda\leq 1/2$, i.e.,
$F(\cdot)$ is a bounded and nondecreasing function on $[-1/2,1/2]$
satisfying
\begin{equation}
  \E{(H_{k+m}-d)\conj{(H_k-d)}} = 
  \int_{-1/2}^{1/2} e^{\ii 2 \pi m \lambda}\d F(\lambda), \quad
  \bigl( k\in\Integers,\;m\in\Integers\bigr),
\end{equation}
where $\ii = \sqrt{-1}$, and where $\conj{A}$ denotes the complex
conjugate of $A$ \cite[p.~474, Thm.~3.2]{doob90}. Since $F(\cdot)$ is
monotonic, it is almost everywhere differentiable, and we denote its
derivative by $F'(\cdot)$. (At the discontinuity points of $F(\cdot)$
the derivative $F'(\cdot)$ is undefined.) For example, if the fading
process $\{H_k,\,k\in\Integers\}$ is IID, then
\begin{equation*}
  F'(\lambda) = 1, \qquad -\frac{1}{2}\leq\lambda\leq\frac{1}{2}.
\end{equation*}

\section{Channel Capacity and the Pre-Log}
\label{sec:capacity}
Channel capacity is defined as the supremum of all achievable
rates. (We refer to \cite[Ch.~8]{coverthomas91} for a definition of an
achievable rate and for a more detailed discussion of channel capacity.)
It was shown (e.g., \cite[Thm.~2]{kim08}) that the
capacity of our channel \eqref{eq:channel} under a peak-power
constraint $\const{A}^2$ on the inputs is given by
\begin{equation}
  \label{eq:capacity}
  C(\SNR) = \lim_{n \to \infty}\frac{1}{n} \sup I(X_1^n;Y_1^n);
\end{equation}
where $\SNR$ is defined as
\begin{equation}
  \label{eq:snr}
  \SNR \triangleq \frac{\const{A}^2}{\sigma^2};
\end{equation}
$A_m^n$ denotes the sequence $A_m,\ldots,A_n$; and where the
maximization is over all joint distributions on $X_1,\ldots,X_n$
satisfying with probability one
\begin{equation}
  \label{eq:power}
  |X_k|^2 \leq \const{A}^2, \qquad k=1,\ldots,n.
\end{equation}
The capacity pre-log is defined as \cite{lapidoth05}
\begin{equation}
  \label{eq:prelogdef}
  \Pi \triangleq \varlimsup_{\SNR \to \infty}
  \frac{C(\SNR)}{\log \SNR}.
\end{equation}

For \emph{Gaussian fading}, i.e., when $\{H_{k}-d,\,k\in\Integers\}$ is a
circularly-symmetric, complex Gaussian
process, the pre-log $\PiG$ is given by the Lebesgue measure of the
set of harmonics where the derivative of the spectral distribution
function is zero, i.e.,
\begin{equation}
  \label{eq:gauss}
  \PiG = \mu\left(\left\{\lambda\colon F'(\lambda)=0\right\}\right),
\end{equation}
where $\mu(\cdot)$ denotes the Lebesgue measure on the interval
$[-1/2,1/2]$; see \cite{lapidoth05}, \cite{lapidoth03_2}. (Here the
subscript ``G'' stands for ``Gaussian''.)

This result indicates that if the fading process is Gaussian and
satisfies
\begin{equation*}
  \mu\left(\left\{\lambda\colon F'(\lambda)=0\right\}\right) > 0,
\end{equation*}
then the corresponding channel capacity grows
logarithmically in the SNR. Note that otherwise the capacity can
increase with the SNR in various ways. For instance, in
\cite{lapidothmoser03_3} fading channels are studied that result in a
capacity which increases double-logarithmically with the SNR, and in
\cite{lapidoth05} spectral distribution functions are presented for
which capacity grows as a fractional power of the logarithm of the
SNR.

\section{Main Result}
\label{sec:results}
We show that, among all stationary \& ergodic fading
processes of a given spectral distribution function and whose law has
no mass point at zero, the Gaussian process gives rise to the smallest
pre-log. This is made precise in the following theorem.

\begin{theorem}
  \label{thm:prelog}
  Consider a mean-$d$, unit-variance, stationary \& ergodic fading
  process $\{H_k,\,k\in\Integers\}$ whose spectral distribution
  function is given by $F(\cdot)$ and whose law satisfies
  \begin{equation*}
    \Prv{H_k=0} = 0, \qquad k \in \Integers.
  \end{equation*}
  Then the corresponding capacity pre-log $\Pi$ is lower bounded by
  \begin{equation}
    \label{eq:mainlb}
    \Pi \geq \mu\left(\left\{\lambda\colon F'(\lambda)=0\right\}\right).
  \end{equation}
\end{theorem}
\begin{proof}
  See Section~\ref{sub:proofprelog}.
\end{proof}

The assumption that the law of the fading process has no mass point
at zero is essential in the following sense. 
\begin{note}
  \label{note:zero}
  There exists a mean-$d$, unit-variance, stationary \& ergodic fading
  process $\{H_k,\,k\in\Integers\}$ of some spectral distribution function
  $F(\cdot)$ such that
  \begin{equation}
    \label{eq:mainup}
    \Pi <
    \mu\left(\left\{\lambda\colon F'(\lambda)=0\right\}\right).
  \end{equation}
  By Theorem~\ref{thm:prelog}, this process must satisfy
  \begin{equation*}
    \Prv{H_k=0} > 0, \qquad k\in\Integers.
  \end{equation*}
\end{note}
\begin{proof}
  See Section~\ref{sub:zero}.
\end{proof}

\begin{note}
  \label{note:phasenoise}
  The inequality in \eqref{eq:mainlb} can be strict. For example,
  consider the \emph{phase-noise channel} with memoryless phase
  noise. This channel can be viewed as a fading channel where the
  fading process $\{H_k,\,k\in\Integers\}$ is given by
  \begin{equation*}
    H_k = e^{\ii \Theta_k}, \qquad k\in\Integers,
  \end{equation*}
  and where $\{\Theta_k,\,k\in\Integers\}$ is IID with $\Theta_k$ being
  uniformly distributed over $[-\pi,\pi)$. This process gives rise to
  a pre-log $\Pi=1/2$, whereas the Gaussian fading of equal spectral
  distribution function yields $\PiG=0$.
\end{note}
\begin{proof}
  For a derivation of the capacity pre-log of the phase-noise channel
  see Section~\ref{sub:phasenoisePP}.
\end{proof}

\section{Proofs}
\label{sec:proofprelog}
This section provides the proofs of our main results. For a proof of
Theorem~\ref{thm:prelog} see Section~\ref{sub:proofprelog}, for a
proof of Note~\ref{note:zero} see Section~\ref{sub:zero}, and for a
proof of Note~\ref{note:phasenoise} see Section~\ref{sub:phasenoisePP}.

\subsection{Proof of Theorem~\ref{thm:prelog}}
\label{sub:proofprelog}
To prove Theorem~\ref{thm:prelog}, we derive in
Section~\ref{sub:lb} a lower bound on the capacity, and proceed in
Section~\ref{sub:prelog} to analyze its asymptotic growth as the SNR
tends to infinity.

\subsubsection{Capacity Lower Bound}
\label{sub:lb}
To derive a lower bound on the capacity we consider inputs
$\{X_k,\,k\in\Integers\}$ that are IID, zero-mean, circularly-symmetric, and for which
$|X_k|^2$ is uniformly distributed over the interval
$\left[0,\const{A}^2\right]$. Our derivation is based on the lower bound
\begin{equation}
  \label{eq:1}
  \frac{1}{n} I(X_1^n;Y_1^n) \geq \frac{1}{n} I(X_1^n;Y_1^n|H_1^n) -
  \frac{1}{n} I(H_1^n;Y_1^n|X_1^n),
\end{equation}
which follows from the chain rule
\begin{IEEEeqnarray}{lCl}
  I(X_1^n;Y_1^n) & = & I(X_1^n,H_1^n;Y_1^n) -
  I(H_1^n;Y_1^n|X_1^n)\nonumber\\
  & = & I(H_1^n;Y_1^n) + I(X_1^n;Y_1^n|H_1^n) - I(H_1^n;Y_1^n|X_1^n)
\end{IEEEeqnarray}
and the nonnegativity of mutual information.

We first study the first term on the right-hand side (RHS) of
\eqref{eq:1}. Making use of the stationarity of the channel and of
the fact that the inputs are IID we have
\begin{equation}
  \label{eq:coherent2}
  \frac{1}{n} I(X_1^n;Y_1^n|H_1^n) = I(X_1;Y_1|H_1). 
\end{equation}
We lower bound the RHS of \eqref{eq:coherent2} as follows. For any
fixed $\Upsilon > 0$
\begin{IEEEeqnarray}{lCl}
  I(X_1;Y_1|H_1) & = & h(H_1X_1+Z_1|H_1)-h(Z_1) \nonumber\\
  & = & \int_{|h_1| \geq \Upsilon} h(H_1X_1+Z_1|H_1=h_1) \d P_{H_1}(h_1) \nonumber\\
  & & {} +\int_{|h_1| < \Upsilon} h(H_1X_1+Z_1|H_1=h_1) \d
  P_{H_1}(h_1) - h(Z_1) \nonumber\\
  & \geq & \int_{|h_1| \geq \Upsilon} h(H_1X_1+Z_1|H_1=h_1) \d
  P_{H_1}(h_1) + \Prv{|H_1|<\Upsilon} h(Z_1)-h(Z_1) \nonumber\\
  & \geq & \int_{|h_1|\geq \Upsilon}\left(\log
  |h_1|^2+h(X_1)\right)\d P_{H_1}(h_1) +\Prv{|H_1|<\Upsilon} h(Z_1)-h(Z_1) \nonumber\\
  & \geq & \Prv{|H_1| \geq \Upsilon}\left(\log\Upsilon^2+h(X_1)\right)
  + \Prv{|H_1|<\Upsilon} h(Z_1) - h(Z_1) \nonumber\\
  & = & \Prv{|H_1| \geq
    \Upsilon}\left(\log\Upsilon^2+\log\pi+h(|X_1|^2)\right) + \Prv{|H_1|<\Upsilon} h(Z_1) - h(Z_1) \nonumber\\
  & = & \Prv{|H_1| \geq \Upsilon}\log \const{A}^2 +
  \Prv{|H_1|\geq\Upsilon}\log\left(\pi\Upsilon^2\right) + \Prv{|H_1|<\Upsilon}
  h(Z_1) - h(Z_1)\nonumber\\
  & = & \Prv{|H_1| \geq \Upsilon}\log \const{A}^2 +
  \Prv{|H_1|\geq\Upsilon}\log\left(\pi\Upsilon^2\right) + \left(\Prv{|H_1|<\Upsilon}-1\right)\log(\pi e\sigma^2) \nonumber\\
  & = &  \Prv{|H_1| \geq \Upsilon}\log \SNR -
  \Prv{|H_1|\geq\Upsilon}\left(1-\log\Upsilon^2\right),
  \label{eq:coherent}
\end{IEEEeqnarray}
where $P_{H_1}(\cdot)$ denotes the distribution function of the fading $H_1$.
Here the third step follows by conditioning the entropy in the
second integral on $X_1$; the fourth step follows by
conditioning the entropy in the first integral on $Z_1$ and by the behavior of differential
entropy under scaling \cite[Thm.~9.6.4]{coverthomas91}; the fifth step follows because over the
range of integration $|h_1|\geq \Upsilon$ we have $\log|h_1|^2 \geq
\log\Upsilon^2$; the sixth step follows because $X_1$ is circularly-symmetric \cite[Lemma
6.16]{lapidothmoser03_3}; the seventh step follows by computing the entropy
of a random variable that is uniformly distributed over the interval
$\left[0,\const{A}^2\right]$; the eighth step follows by
evaluating the entropy of a zero-mean, variance-$\sigma^2$,
circularly-symmetric, complex Gaussian random variable $h(Z_k)=\log(\pi
e\sigma^2)$; and the last step follows from $\Prv{|H_1| \geq
  \Upsilon} = 1 - \Prv{|H_1|<\Upsilon}$.

We next turn to the second term on the RHS of \eqref{eq:1}. In order
to upper bound it we proceed along the lines of
\cite{denghaimovich04}, but for non-Gaussian fading. Let $\vect{Y}$, $\vect{H}$, and $\vect{Z}$ be
the random vectors $\trans{(Y_1,\ldots,Y_n)}$,
$\trans{(H_1,\ldots,H_n)}$, and $\trans{(Z_1,\ldots,Z_n)}$ (where
$\trans{\vect{A}}$ denotes the transpose of $\vect{A}$), and let
$\mat{X}$ be a diagonal matrix with diagonal entries
$x_1,\ldots,x_n$. It follows from \eqref{eq:channel} that
\begin{equation}
  \vect{Y} = \mat{X} \vect{H} + \vect{Z}.
\end{equation}
The conditional covariance matrix of $\vect{Y}$,
conditional on $x_1,\ldots,x_n$, is given by
\begin{equation}
  \Econd{\left(\vect{Y}-\E{\vect{Y}}\right)\hermi{\left(\vect{Y}-\E{\vect{Y}}\right)}}{X_1^n=x_1^n} 
  = \mat{X} \mat{K}_{\vect{H}\vect{H}}\hermi{\mat{X}} + \sigma^2 \mat{I}_n,
\end{equation}
where $\mat{I}_n$ is the $n \times n$ identity matrix,
$\hermi{(\cdot)}$ denotes Hermitian conjugation, and 
\begin{equation}
  \mat{K}_{\vect{H}\vect{H}} \triangleq \E{(\vect{H}-\E{\vect{H}})\hermi{(\vect{H}-\E{\vect{H}})}}.
\end{equation}
Let $\det \mat{A}$ denote the determinant of the matrix $\mat{A}$. Using the entropy maximizing
property of circularly-symmetric Gaussian vectors \cite[Thm.~9.6.5]{coverthomas91}, we have
\begin{IEEEeqnarray}{lCl}
  \frac{1}{n} I(H_1^n;Y_1^n|X_1^n) & = & \frac{1}{n}
  h(Y_1^n|X_1^n)-\frac{1}{n} h(Z_1^n) \nonumber\\
  & \leq &
  \frac{1}{n} \E{\log\det\left(\mat{I}_n+\frac{1}{\sigma^2}\vmat{X}\mat{K}_{\vect{H}\vect{H}}\hermi{\vmat{X}}\right)}\nonumber\\
  & = &
  \frac{1}{n} \E{\log\det\left(\mat{I}_n+\frac{1}{\sigma^2}\mat{K}_{\vect{H}\vect{H}}\hermi{\vmat{X}}\vmat{X}\right)}\nonumber\\
  & \leq &
  \frac{1}{n} \log\det\left(\mat{I}_n+\frac{\const{A}^2}{\sigma^2}\mat{K}_{\vect{H}\vect{H}}\right)
  \nonumber\\
  & = &
  \frac{1}{n} \log\det\left(\mat{I}_n+\SNR\,\mat{K}_{\vect{H}\vect{H}}\right)\nonumber\\
  & = & \frac{1}{n} \sum_{k=1}^n \log(1+\SNR\,\lambda_k),
  \label{eq:noncoherent}
\end{IEEEeqnarray}
where $\vmat{X}$ is a random diagonal matrix with diagonal entries
$X_1,\ldots,X_n$, and where $\lambda_1,\ldots,\lambda_n$ denote the
eigenvalues of $\mat{K}_{\vect{H}\vect{H}}$.
Here the third step follows from the identity
$\det(\mat{I}_n+\mat{A}\mat{B})= \det(\mat{I}_n+\mat{B}\mat{A})$; the
fourth step follows from \eqref{eq:power} which implies that $\const{A}^2
\mat{I}_n - \hermi{\vmat{X}}\vmat{X}$ is positive semidefinite with
probability one; the fifth step follows from the definition of
$\SNR$ \eqref{eq:snr}; and the last step follows because the
determinant of a matrix is given by the product of its eigenvalues. 

To evaluate the RHS of \eqref{eq:noncoherent} in the limit as $n$
tends to infinity, we apply Szeg\"o's Theorem on the asymptotic
behavior of the eigenvalues of Hermitian Toeplitz matrices
\cite{grenanderszego58} (see also \cite[Thm.~2.7.13]{simon05}). We
obtain
\begin{IEEEeqnarray}{lCl}
  \lim_{n \to \infty} \frac{1}{n}
  I(H_1^n;Y_1^n|X_1^n) & \leq & \lim_{n\to\infty} \frac{1}{n}
  \sum_{k=1}^n \log(1+\SNR\,\lambda_k) \nonumber\\
  & = & \int_{-1/2}^{1/2} \log\bigl(1+\SNR\,F'(\lambda)\bigr) \d\lambda.\label{eq:spectrum}
\end{IEEEeqnarray}

Combining \eqref{eq:1}, \eqref{eq:coherent2}, \eqref{eq:coherent}, and
\eqref{eq:spectrum} yields the final lower bound
\begin{IEEEeqnarray}{lCl}
  C(\SNR) & \geq &  \Prv{|H_1| \geq \Upsilon}\log \SNR - \Prv{|H_1|\geq
    \Upsilon}\left(1-\log\Upsilon^2\right) \nonumber\\
  & & {} - \int_{-1/2}^{1/2} \log\left(1+\SNR
  F'(\lambda)\right)\d\lambda, \qquad\qquad\qquad\qquad\quad \SNR>0, \label{eq:LB}
\end{IEEEeqnarray}
which holds for any fixed $\Upsilon>0$. Note that this lower bound
applies to all mean-$d$, unit-variance, stationary \& ergodic fading
processes $\{H_k,\,k\in\Integers\}$ with spectral distribution
function $F(\cdot)$.

\subsubsection{Asymptotic Analysis}
\label{sub:prelog}
In the following we prove \eqref{eq:mainlb} by computing the
limiting ratio of the lower bound \eqref{eq:LB} to $\log\SNR$ as
$\SNR$ tends to infinity.

We first show that
\begin{IEEEeqnarray}{lCl}
  \lim_{\SNR \to \infty} \int_{-1/2}^{1/2}
  \frac{\log\bigl(1+\SNR\,F'(\lambda)\bigr)}{\log\SNR}\d\lambda & = &
  \mu\left(\left\{\lambda\colon F'(\lambda)>0\right\}\right).\label{eq:all}
\end{IEEEeqnarray}
To this end, we divide the integral into three parts, depending
on whether $\lambda$ takes part in the set $\set{S}_1$, $\set{S}_2$,
or $\set{S}_3$, where
\begin{IEEEeqnarray}{lCl}
  \set{S}_1 & \triangleq & \{\lambda \in [-1/2,1/2]\colon F'(\lambda)=0\}\\
  \set{S}_2 & \triangleq & \{\lambda \in [-1/2,1/2]\colon F'(\lambda)\geq
  1\}\\
  \set{S}_3 & \triangleq & \{\lambda \in [-1/2,1/2]\colon 0<F'(\lambda)<1\}.
\end{IEEEeqnarray}
 For $\lambda \in \set{S}_1$ the integrand is zero and hence
\begin{equation}
  \lim_{\SNR\to\infty} \int_{\set{S}_1}\frac{\log\bigl(1+\SNR\,
      F'(\lambda)\bigr)}{\log\SNR}\d\lambda = 0.\label{eq:zero}
\end{equation}
For $\lambda \in \set{S}_2$, i.e., when $F'(\lambda) \geq 1$, we note
that for sufficiently large SNR the function
\begin{equation*}
  \SNR \mapsto \frac{\log\bigl(1+\SNR\,F'(\lambda)\bigr)}{\log\SNR}
\end{equation*}
is monotonically decreasing in $\SNR$. Therefore, applying the
Monotone Convergence Theorem \cite[Thm.~1.26]{rudin87}, we have
\begin{IEEEeqnarray}{lCl}
  \lim_{\SNR\to\infty}
    \int_{\set{S}_2}\frac{\log\bigl(1+\SNR\,F'(\lambda)\bigr)}{\log\SNR}\d\lambda & = &  \int_{\set{S}_2} \lim_{\SNR\to\infty}\frac{\log\bigl(1+\SNR
        \,F'(\lambda)\bigr)}{\log\SNR}\d\lambda \nonumber\\
  & = & \mu\left(\set{S}_2\right)\nonumber\\
  & = & \mu\left(\left\{\lambda\colon F'(\lambda)\geq 1\right\}\right).\label{eq:monotone}
\end{IEEEeqnarray}
For $\lambda \in \set{S}_3$, i.e., when $0<F'(\lambda)<1$, we have
\begin{equation}
  0 < \frac{\log\bigl(1+\SNR\,
      F'(\lambda)\bigr)}{\log\SNR} < \frac{\log(1+\SNR)}{\log\SNR} \leq
      \log(1+e), \quad \SNR \geq e,
\end{equation}
where the last step follows because, for sufficiently large
SNR, the function
\begin{equation*}
\SNR \mapsto \frac{\log(1+\SNR)}{\log\SNR}
\end{equation*}
is monotonically decreasing in $\SNR$. Since $\log(1+e)$ is integrable
over $\set{S}_3$, we can apply the Dominated Convergence Theorem
\cite[Thm.~1.34]{rudin87} to obtain
\begin{IEEEeqnarray}{lCl}
  \lim_{\SNR\to\infty}
    \int_{\set{S}_3}\frac{\log\bigl(1+\SNR\,
        F'(\lambda)\bigr)}{\log\SNR}\d\lambda & = & \int_{\set{S}_3} \lim_{\SNR\to\infty}\frac{\log\bigl(1+\SNR\,
      F'(\lambda)\bigr)}{\log\SNR}\d\lambda \nonumber\\
  & = & \mu\left(\set{S}_3\right) \nonumber\\
  & = & \mu\left(\left\{\lambda\colon 0<F'(\lambda)<1\right\}\right).\label{eq:dominated}
\end{IEEEeqnarray}
Adding \eqref{eq:zero}, \eqref{eq:monotone}, and
\eqref{eq:dominated} yields \eqref{eq:all}.

To continue with the asymptotic analysis of \eqref{eq:LB} we note
that by \eqref{eq:all}
\begin{IEEEeqnarray}{lCl}
  \Pi & \triangleq &
  \varlimsup_{\SNR\to\infty}\frac{C(\SNR)}{\log\SNR}\nonumber\\
  & \geq & \Prv{|H_1|\geq \Upsilon} -
  \mu\left(\left\{\lambda:F'(\lambda)>0\right\}\right)\nonumber\\
  & = & \mu\left(\left\{\lambda\colon F'(\lambda)=0\right\}\right)-\Prv{|H_1|<\Upsilon}\label{eq:withgamma}
\end{IEEEeqnarray}
for any $\Upsilon>0$. If the law of the fading
process has no mass point at zero, then
\begin{equation}
  \lim_{\Upsilon \downarrow 0}\Prv{|H_1|<\Upsilon} =0,
\end{equation}
and \eqref{eq:mainlb} therefore follows from \eqref{eq:withgamma} by
letting $\Upsilon$ tend to zero from above.

\subsection{Proof of Note~\ref{note:zero}}
\label{sub:zero}
We prove Note~\ref{note:zero} by demonstrating that there exists a
stationary \& ergodic fading process of some spectral distribution
function $F(\cdot)$ for which
\begin{equation*}
  \Pi < \mu\left(\{\lambda\colon F'(\lambda)=0\}\right).
\end{equation*}
By Theorem~\ref{thm:prelog}, the law of such a process must have a
mass point at zero, i.e.,
\begin{equation*}
  \Prv{H_k=0} > 0, \qquad k\in\Integers.
\end{equation*}
To this end, we first show that the capacity pre-log is
upper bounded by
\begin{equation}
  \label{eq:zeroub}
  \Pi \leq \Prv{|H_1|>0}.
\end{equation}
Indeed, the capacity $C(\SNR)$ does not decrease when
the receiver additionally knows the realization of
$\{H_k,\,k\in\Integers\}$, and when the inputs have to satisfy an
average-power constraint rather than a peak-power constraint, i.e.,
\begin{equation}
  \label{eq:CSI1}
  C(\SNR) \leq \lim_{n\to\infty} \frac{1}{n} \sup I(X_1^n;Y_1^n|H_1^n),
\end{equation}
where the maximization is over all input distributions on
$X_1,\ldots,X_n$ satisfying the average-power constraint
\begin{equation}
  \label{eq:avgpower}
  \frac{1}{n} \sum_{k=1}^n \frac{\E{|X_k|^2}}{\sigma^2} \leq \SNR.
\end{equation}
(This follows because the availability of additional information
cannot decrease the capacity, and because any distribution on the
inputs satisfying the peak-power constraint \eqref{eq:power} satisfies
also \eqref{eq:avgpower}.) It is well known that the expression on the
RHS of \eqref{eq:CSI1} is equal to
\begin{equation}
  \lim_{n\to\infty} \frac{1}{n} \sup I(X_1^n;Y_1^n|H_1^n) = \E{\log(1+|H_1|^2\,\SNR)}
\end{equation}
(e.g., \cite[eq.~(3.3.10)]{biglieriproakisshamai98}), which
can be further upper bounded by
\begin{IEEEeqnarray}{lCl}
  \E{\log(1+|H_1|^2\,\SNR)} & = &
  \Prv{|H_1|>0}\,\Econd{\log(1+|H_1|^2\,\SNR)}{|H_1|>0}\nonumber\\
  & \leq &
  \Prv{|H_1|>0}\,\log\bigl(1+\Econd{|H_1|^2}{|H_1|>0}\,\SNR\bigr)\nonumber\\
  & = & \Prv{|H_1|>0}\,\log\biggl(1+\frac{\SNR}{\Prv{|H_1|>0}}\biggr).\label{eq:CSI2}
\end{IEEEeqnarray}
Here the first step follows by writing the expectation as
\begin{IEEEeqnarray*}{lCl}
  \E{\log(1+|H_1|^2\,\SNR)} & = &
  \Prv{|H_1|=0}\,\Econd{\log(1+|H_1|^2\,\SNR)}{|H_1|=0} \\
  & & {} + \Prv{|H_1|>0}\,\Econd{\log(1+|H_1|^2\,\SNR)}{|H_1|>0},
\end{IEEEeqnarray*}
and by noting then that $\Econd{\log(1+|H_1|^2\,\SNR)}{|H_1|=0}=0$; the
second step follows from Jensen's inequality; and the last
step follows because $\E{|H_1|^2}=1$, which implies
\begin{equation*}
  \Econd{|H_1|^2}{|H_1|>0} = \frac{1}{\Prv{|H_1|>0}}.
\end{equation*}
Dividing the RHS of \eqref{eq:CSI2} by $\log\SNR$, and computing the
limit as $\SNR$ tends to infinity yields \eqref{eq:zeroub}.

In view of \eqref{eq:zeroub}, it suffices to demonstrate that there
exists a fading process of some spectral distribution function
$F(\cdot)$ that satisfies
\begin{equation}
  \label{eq:zerocond}
  \Prv{|H_1|>0} < \mu\left(\{\lambda:F'(\lambda)=0\}\right).
\end{equation}

A first attempt of defining such a process (which, alas, does not
work) is
\begin{equation}
  \ldots,H_{-1},H_0,H_1,H_2,\ldots = \left\{ \begin{array}{ll} \ldots,0,0,0,0,\ldots \quad &
  \text{with probability $\delta$}\\[10pt] \ldots,B_{-1},B_0,B_1,B_2,\ldots \quad &
  \text{with probability $1-\delta$,}\end{array}\right.
\end{equation}
where $\{B_k,\,k\in\Integers\}$ is a zero-mean, circularly-symmetric,
stationary \& ergodic, complex Gaussian process of variance
$1/(1-\delta)$ and of spectral distribution function $G(\cdot)$; and
where $\delta$ and $G(\cdot)$ are chosen so that
\begin{equation}
  1-\delta < \mu\left(\{\lambda:G'(\lambda)=0\}\right).
\end{equation}
This process satisfies \eqref{eq:zerocond} because
$\Prv{|H_1|>0}=1-\delta$, and because
\begin{equation}
  \E{(H_{k+m}-d)\conj{(H_k-d)}} = (1-\delta)\,
  \E{B_{k+m}\conj{B}_k},
\end{equation}
which implies that $F(\lambda)=(1-\delta)G(\lambda)$ almost
everywhere, so
\begin{equation}
  \mu\left(\{\lambda:F'(\lambda)=0\}\right) =
  \mu\left(\{\lambda:G'(\lambda)=0\}\right).
\end{equation}
Alas, the above fading process is stationary but not ergodic.

In the following, we exhibit a fading process that is stationary \&
ergodic and satisfies \eqref{eq:zerocond}. Let
\begin{equation}
  \ldots,A_{-1},A_0,A_1,A_2,\ldots = \left\{\begin{array}{ll} \ldots,0,1,0,1,\ldots
  \quad & \text{with probability $\frac{1}{2}$}\\[10pt] \ldots,1,0,1,0,\ldots \quad &
  \text{with probability $\frac{1}{2}$}, \end{array}\right.
\end{equation}
and let $\{B_k,\,k\in\Integers\}$ be a zero-mean, variance-$2$, circularly-symmetric,
stationary \& ergodic, complex Gaussian process of spectral
distribution function $G(\cdot)$. Furthermore let
$\{A_k,\,k\in\Integers\}$ and $\{B_k,\,k\in\Integers\}$ be independent
of each other. We shall consider fading processes of the form
\begin{equation}
  H_k = A_k \cdot B_k, \qquad k\in\Integers.
\end{equation}
Note that $\{H_k,\,k\in\Integers\}$ is of zero mean, and its law has a
mass point at zero
\begin{equation}
  \label{eq:zerozero}
  \Prv{|H_k|>0} = \Prv{A_k=1} = \frac{1}{2}, \qquad k\in\Integers.
\end{equation}

We first argue that $\{H_k,k\in\Integers\}$ is stationary \&
ergodic. Indeed, $\{A_k,\,k\in\Integers\}$ is stationary \&
ergodic. And since a Gaussian process is ergodic if, and only if, it
is weakly-mixing (see, e.g., \cite[Sec.~II]{sethuramanhajek05}),
we have that $\{B_k,\,k\in\Integers\}$ is stationary \&
weakly-mixing. (See \cite[Sec.~2.6]{petersen83} for a definition of
weakly-mixing stochastic processes.) It thus follows from \cite[Prop.~1.6]{brown76} that the process
$\{(A_k,B_k),\,k\in\Integers\}$ is jointly stationary \& ergodic,
which implies that $\{H_k,\,k\in\Integers\} = \{A_k\cdot
B_k,\,k\in\Integers\}$ is stationary \& ergodic.

We next demonstrate that $G(\cdot)$ can be chosen so that
$\{H_k,\,k\in\Integers\}$ satisfies \eqref{eq:zerocond}. We choose
\begin{equation}
  G'(\lambda) = \left\{\begin{array}{ll}\displaystyle \frac{1}{\WW}, \quad &
      \text{if $|\lambda|\leq \WW$}\\[10pt] \displaystyle 0,\quad & \text{otherwise}\end{array} \right.
\end{equation}
for some $\WW\in(0,1/8)$, which corresponds to the autocovariance function
\begin{equation*}
  \E{B_{k+m}\conj{B}_k} = 2 \sinc(2\WW m), \qquad m\in\Integers.
\end{equation*}
Here $\sinc(\cdot)$ denotes the sinc-function, i.e.,
$\sinc(x)=\sin(\pi x)/(\pi x)$ for $|x|>0$ and
$\sinc(0)=1$. Using that
\begin{equation*}
\E{A_{k+m}\conj{A}_{k}}=\frac{1}{2}\I{\text{$m$ is even}}, \quad m\in\Integers
\end{equation*}
(where $\I{\text{statement}}$ is $1$ if the statement is true, and $0$
otherwise), we have for the autocovariance function of $\{H_k,\,k\in\Integers\}$
\begin{IEEEeqnarray}{lCl}
  \E{H_{k+m}\conj{H}_k} & = & \E{A_{k+m}B_{k+m}\conj{A}_k
    \conj{B}_k}\nonumber\\
  & = & \E{A_{k+m} \conj{A}_{k}}\E{B_{k+m}\conj{B}_k}\nonumber\\
  & = & \I{\text{$m$ is even}}\cdot\sinc(2\WW m), \quad m\in\Integers,
\end{IEEEeqnarray}
and the corresponding spectrum is given by
\begin{equation}
  F'(\lambda) = \left\{\begin{array}{ll} \frac{1}{4\WW},\quad &
  \text{if $|\lambda|\leq\WW$ or $\frac{1}{2}-\WW\leq |\lambda| \leq
  \frac{1}{2}$}\\[10pt] 0, & \text{otherwise.}\end{array}\right.
\end{equation}
Evaluating the Lebesgue measure of the set of harmonics where
$F'(\lambda)=0$, we have
\begin{equation}
  \mu\left(\{\lambda\colon F'(\lambda)=0\}\right) = 1-4\WW,
\end{equation}
and it follows from \eqref{eq:zerozero} that
\begin{equation*}
  \Prv{|H_k|>0} = \frac{1}{2} <
  \mu\left(\{\lambda\colon F'(\lambda)=0\}\right), \quad \text{for $\WW<\frac{1}{8}$}.
\end{equation*}
Thus there exist stationary \& ergodic fading
processes whose law has a mass point at zero and that give rise to a
capacity pre-log that is strictly smaller than the pre-log of a
Gaussian fading channel of equal spectral distribution function.

\subsection{Proof of Note~\ref{note:phasenoise}}
\label{sub:phasenoisePP}
To prove Note~\ref{note:phasenoise}, we first notice that, since the
phase noise is memoryless, the derivative of the spectral distribution
function is
\begin{equation*}
  F'(\lambda) = 1, \qquad -\frac{1}{2}\leq\lambda\leq\frac{1}{2}.
\end{equation*}
Hence the capacity pre-log of the Gaussian fading channel of spectral
distribution function $F(\cdot)$ equals
\begin{equation}
  \PiG = \mu\left(\{\lambda\colon F'(\lambda) = 0\}\right) = 0.
\end{equation}

It thus remains to show that the pre-log of the phase-noise channel with
memoryless phase noise is equal to
\begin{equation}
  \label{eq:prelogphasenoise}
  \Pi = \frac{1}{2}.
\end{equation}
In \cite{lapidoth02_2} it was shown that at high SNR the capacity of
the phase-noise channel under an average-power constraint on the
inputs is given by
\begin{equation}
  C_{\text{Avg}}(\SNR) = \frac{1}{2}\log\biggl(1+\frac{\SNR}{2}\biggr)+ o(1),
\end{equation}
where $o(1)$ tends to zero as $\SNR$ tends to zero. (The subscript
``Avg'' indicates that the inputs satisfy an average-power
constraint and not a peak-power constraint.) Since any distribution on
the inputs satisfying the peak-power constraint \eqref{eq:power}
satisfies also the average-power constraint, it follows that
$C(\SNR)\leq C_{\text{Avg}}(\SNR)$ and hence
\begin{equation}
  \label{eq:appprelogavg}
  \Pi \leq \frac{1}{2}.
\end{equation}
To prove \eqref{eq:prelogphasenoise} it thus suffices to show that
$\Pi\geq\frac{1}{2}$. To this end, we first note that, since the phase
noise is memoryless, we have
\begin{equation}
  \label{eq:memorylessC}
  C(\SNR) = \sup I(X_1;Y_1),
\end{equation}
where the maximization is over all distributions on $X_1$ satisfying
with probability one
\begin{equation*}
  |X_1| \leq \const{A}.
\end{equation*}

We derive a lower bound on $C(\SNR)$ by evaluating the
RHS of \eqref{eq:memorylessC} for $X_1$ being a zero-mean,
circularly-symmetric, complex random variable with $|X_1|^2$ uniformly
distributed over the interval $\bigl[0,\const{A}^2\bigr]$. We have
\begin{IEEEeqnarray}{lCl}
  I(X_1;Y_1) & \geq & I\bigl(X_1;|Y_1|^2\bigr) \nonumber\\
  & = & h\bigl(|Y_1|^2\bigr) - h\bigl(|Y_1|^2\bigm|X_1\bigr)
  \nonumber\\
  & \geq & h\bigl(|X_1|^2\bigr) - h\bigl(|Y_1|^2\bigm|X_1\bigr),\label{eq:app1}
\end{IEEEeqnarray}
where the first step follows from the data processing inequality
\cite[Thm.~2.8.1]{coverthomas91}; and the last step follows
by the circular symmetry of $X_1$ \cite[p.~3, after
eq.~(20)]{lapidoth02_2}.

Computing the differential entropy of a uniformly distributed random
variable, the first term on the RHS of \eqref{eq:app1} becomes
\begin{equation}
  \label{eq:app2}
  h\bigl(|X_1|^2\bigr) = \log\const{A}^2.
\end{equation}
As to the second term, we note that, for a given $X_1=x_1$, the random
variable $2/\sigma^2\,|Y_1|^2$ has a noncentral chi-square
distribution with noncentrality parameter $2/\sigma^2\,|x_1|^2$ and
two degrees of freedom. Its differential entropy can be upper bounded
by \cite[eq.~(8)]{lapidoth02_2}
\begin{IEEEeqnarray}{lCl}
  h\bigl(|Y_1|^2\bigm|X_1\bigr) & \leq & \frac{1}{2} \E{\log\biggl(4\pi
  e\Bigl(2+2\frac{2}{\sigma^2}|X_1|^2\Bigr)\biggr)} -
  \log\frac{2}{\sigma^2} \nonumber\\
  & \leq & \frac{1}{2}\log\biggl(4\pi e\Bigl(2+2\frac{2}{\sigma^2}\const{A}^2\Bigr)\biggr)-
  \log\frac{2}{\sigma^2}, \label{eq:app3}
\end{IEEEeqnarray}
where the last step follows because $|X_1|\leq\const{A}$ with
probability one. Combining \eqref{eq:app2} and \eqref{eq:app3} with
\eqref{eq:app1} yields thus
\begin{equation}
  \label{eq:applast}
  I(X_1;Y_1) \geq \frac{1}{2} \log\SNR + o(\log\SNR),
\end{equation}
where
\begin{equation*}
  \lim_{\SNR\to\infty} \frac{o(\log\SNR)}{\log\SNR} = 0.
\end{equation*}
We finally obtain the lower bound
\begin{equation*}
  \Pi \geq \frac{1}{2}
\end{equation*}
upon dividing the RHS of \eqref{eq:applast} by $\log\SNR$ and letting then $\SNR$
tend to infinity.

\section{Extension to MISO Fading Channels}
\label{sec:MISO}
Theorem~\ref{thm:prelog} can be extended to multiple-input
single-output (MISO) fading channels with memory, when the fading
processes corresponding to the different transmit antennas are
independent. For such channels, the channel
output $Y_k\in\Complex$ at time $k\in\Integers$ corresponding to the
channel input $\vect{x}_k\in\Complex^{\nt}$ (where $\nt$ stands for
the number of antennas at the transmitter) is given by
\begin{equation}
  \label{eq:MISOchannel}
  Y_k = \trans{\vect{H}}_k \vect{x}_k + Z_k, \qquad k \in\Integers,
\end{equation}
where $\vect{H}_k=\trans{\Bigl(H_k^{(1)},\ldots,H_k^{(\nt)}\Bigr)}$,
and where the processes
\begin{equation*}
\bigl\{H_k^{(1)},\,k\in\Integers\bigr\},\bigl\{H_k^{(2)},\,k\in\Integers\bigr\}\ldots,\bigl\{H_k^{(\nt)},\,k\in\Integers\bigr\}
\end{equation*}
are jointly stationary \& ergodic and independent. We assume that for each $t=1,\ldots,\nt$ the process
$\bigl\{H_k^{(t)},\,k\in\Integers\bigr\}$ is of mean $d_t$, of unit variance, and of spectral
distribution function $F_t(\cdot)$.
We further assume that
\begin{equation}
  \Prv{H_k^{(1)}=0}=\Prv{H_k^{(2)}=0}=\ldots=\Prv{H_k^{(\nt)}=0}=0, \quad
  k\in\Integers.
\end{equation}
The additive noise $\{Z_k,\,k\in\Integers\}$ is defined as in
Section~\ref{sec:channel}.

The capacity of this channel is given by \eqref{eq:capacity}, but with
$X_1^n$ replaced by $\vect{X}_1^n$, and with the peak-power constraint
\eqref{eq:power} altered accordingly:
\begin{equation}
  \norm{\vect{X}_k} \leq \const{A} \quad \text{with
  probability one,} \qquad k\in\Integers,
\end{equation}
where $\norm{\vect{a}}$ denotes the Euclidean norm of the vector
$\vect{a}$, i.e.,
\begin{equation}
  \norm{\vect{a}} = \sqrt{\sum_{\ell=1}^{\const{L}} |a_{\ell}|^2}, \qquad
  \vect{a} = \trans{(a_1,\ldots,a_{\const{L}})}.
\end{equation}
Let $\Xi$ denote the pre-log of MISO fading channels. Following
\eqref{eq:prelogdef}, we define $\Xi$ as
\begin{equation}
  \Xi \triangleq \varlimsup_{\SNR\to\infty} \frac{C(\SNR)}{\log\SNR}.
\end{equation}
For Gaussian fading, i.e., when $\{H^{(t)}-d_t,\,k\in\Integers\}$, $1\leq t\leq\nt$ are
circularly-symmetric, complex Gaussian processes, the pre-log was shown to be given by
\cite[Cor.~13]{kochlapidoth05_3}
\begin{equation}
  \Xi_{\text{G}} = \max_{1\leq t\leq\nt} \mu\left(\{\lambda\colon
  F'_t(\lambda)=0\}\right).
\end{equation}
(A proof of this result can be found in \cite[Sec.~7.2.2]{koch04}.)

Proving that the capacity pre-log $\Xi$ of MISO fading channels is
lower bounded by the pre-log of the MISO Gaussian fading channel of
equal spectral distribution functions---namely
$F_1(\cdot),\ldots,F_{\nt}(\cdot)$---is straightforward. Let $\Pi_t$,
$1\leq t\leq\nt$ denote the capacity pre-log of a single-antenna
fading channel with fading process
$\bigl\{H_k^{(t)},\,k\in\Integers\bigr\}$, and let
\begin{equation*}
  t_{\star} = \arg \max_{1\leq t\leq\nt} \Pi_t.
\end{equation*} 
By signaling only from antenna $t_{\star}$ while keeping the others silent,
we can achieve the pre-log $\Pi_{t_{\star}}$, so
\begin{equation}
  \label{eq:MISOlb}
  \Xi \geq \max_{1\leq t\leq\nt} \Pi_t.
\end{equation}
Theorem~\ref{thm:prelog} yields then
\begin{equation}
  \Pi_t \geq \mu\left(\{\lambda\colon F'_t(\lambda)=0\}\right), \qquad 1\leq t\leq\nt,
\end{equation}
which together with \eqref{eq:MISOlb} proves the claim
\begin{equation}
  \Xi \geq \max_{1\leq t\leq\nt}\mu\left(\{\lambda\colon F'_t(\lambda)=0\}\right).
\end{equation}

\section{Summary and Discussion}
\label{sec:discussion}
We showed that, among all stationary \& ergodic fading
processes of a given spectral distribution function and whose law has
no mass point at zero, the Gaussian process gives rise to the smallest
capacity pre-log. We further showed that if the fading law
is allowed to have a mass point at zero, then the above statement is
not necessarily true anymore. Roughly speaking, we can say that for a
large class of fading processes the Gaussian process is the
worst. This demonstrates the robustness of the Gaussian assumption in
the analysis of fading channels at high SNR.

To give an intuition why Gaussian processes give rise to the smallest
pre-log, we recall that for Gaussian fading \cite[eqs.~(33) \& (47)]{lapidoth05}
\begin{equation*}
  C(\SNR) = \log\frac{1}{\eps^2_{\text{pred}}(1/\SNR)} + o(\log\SNR),
\end{equation*}
where $\eps^2_{\text{pred}}(\delta)$ denotes the mean-square error in
predicting the present fading $H_0$ from a variance-$\delta$ noisy
observation of its past $H_{-1}+W_{-1},H_{-2}+W_{-2},\ldots$ (with
$\{W_k,\,k\in\Integers\}$ being a sequence of IID, zero-mean,
variance-$\delta$, circularly-symmetric, complex Gaussian random
variables). Thus for Gaussian fading the capacity pre-log is determined by
$\eps^2_{\text{pred}}(1/\SNR)$, and it is plausible that also the
pre-log of non-Gaussian fading channels is connected with the ability of
predicting the present fading from a noisy observation of its
past. Since, among all stationary \& ergodic processes of a given
spectral distribution function, the Gaussian process is hardest to
predict, it is therefore plausible that the Gaussian process gives
rise to the smallest pre-log.


\end{document}